\documentclass[manuscript]{acmart}

\usepackage{svg} 

\usepackage{multirow}
\usepackage{float}
\usepackage{subfloat} 

\AtBeginDocument{%
  }

\begin{document}


\title{Towards Fair and Rigorous Evaluations: Hyperparameter Optimization for Top-N Recommendation Task with Implicit Feedback}

\author{Hui Fang}
\orcid{0000-0001-9788-6634}
\affiliation{%
  \institution{Shanghai University of Finance and Economics}
  \city{Shanghai}
  \country{China}
}
\email{fang.hui@mail.shufe.edu.cn} 
\thanks{*Corresponding author.}

\author{Xu Feng}
\affiliation{%
  \institution{Shanghai University of Finance and Economics}
  \city{Shanghai}
  \country{China}
}

\author{Lu Qin}
\affiliation{%
  \institution{Shanghai University of Finance and Economics}
  \city{Shanghai}
  \country{China}
}

\author{Zhu Sun}
\orcid{0000-0002-3350-7022}
\affiliation{%
  \institution{Singapore University of Technology and Design}
  \city{Singapore}
  \country{Singapore}
}

\renewcommand{\shortauthors}{Qin et al.}

\begin{abstract}
The widespread use of the internet has led to an overwhelming amount of data, which has resulted in the problem of information overload. Recommender systems have emerged as a solution to this problem by providing personalized recommendations to users based on their preferences and historical data. However, as recommendation models become increasingly complex, finding the best hyperparameter combination for different models has become a challenge. The high-dimensional hyperparameter search space poses numerous challenges for researchers, and failure to disclose hyperparameter settings may impede the reproducibility of research results. In this paper, we investigate the Top-N implicit recommendation problem and focus on optimizing the benchmark recommendation algorithm commonly used in comparative experiments using hyperparameter optimization algorithms. We propose a research methodology that follows the principles of a fair comparison, employing seven types of hyperparameter search algorithms to fine-tune six common recommendation algorithms on three datasets. We have identified the most suitable hyperparameter search algorithms for various recommendation algorithms on different types of datasets as a reference for later study. This study contributes to algorithmic research in recommender systems based on hyperparameter optimization, providing a fair basis for comparison.
\end{abstract}



\maketitle

\section{Introduction}
The internet has revolutionized the way people access information, providing them with unparalleled convenience. However, this convenience has come at a cost since the overwhelming amount of data available has led to the problem of information overload. Recommender system which gives users personalized recommendations based on their preferences and historical data, has thus emerged as a solution to this problem. It has seen significant success in various domains, from e-commerce to social networking and multimedia, over the past two decades \cite{sun2019research}.

However, previous studies incline to design increasingly complex recommendation models with a growing number of hyperparameters. And, the hyperparameter optimization problem is complicated by the fact that the objective function is typically a black box function, due to the tailoring of model structure to specific business needs. As a result, the choice of hyperparameter values significantly affects the model's final performance. 
To conclude,
the challenge of finding the best hyperparameter combination for different models has arisen from two perspectives.

On the one hand,
failure to disclose the settings of all hyperparameters in published experiments may impede the reproducibility of research results. A considerable amount of time in model designs is spent on tuning hyperparameters to achieve the reported performance level. Besides, \cite{pinto2009high} highlighted that in large and multi-layer complex models, the hyperparameter search space is typically vast, and evaluating a particular set of hyperparameters incurs high computational costs. Consequently, a significant portion of the space remains unexplored. In that case, if experimental results do not meet expectations, it is unclear whether the issue is with the model's rationality, or because the optimal model instance has not been found. 

On the other hand, failure to fine-tune the performance of recommendation algorithms to their optimal levels can also raise fairness comparison concerns. For example,~\cite{lin2019neural} demonstrated that via a careful hyperparameter optimization, new neural network-based methods did not significantly outperform existing baseline methods. Coates et al.~\cite{coates2011analysis} opted to use relatively simple algorithms with hyperparameter optimization, rather than innovative modeling or machine learning strategies, to improve performance in image classification problems.~\cite{ludewig2018evaluation} suggested that even recent session-based recommender system algorithms can be replaced by simple methods such as nearest neighbors in most cases.
To solve the fairness problem, recent work~\cite{sun2020we,sun2022daisyrec} implemented a fair comparison framework, i.e., DaisyRec 2.0, by reviewing a large number of papers published in top conferences, which filled the gap of fair and rigorous evaluations for top-N recommendation task from the aspects of dataset partitioning, evaluation metrics, and experimental settings. Our work builds upon this framework to further advance research on automatic hyperparameter optimization, enabling us to identify more suitable baseline recommendation models and matching hyperparameter search algorithms for different scenarios by taking into account the characteristics of different datasets.

\begin{figure}[htbp]
  \centering
  \includegraphics[width=\linewidth]{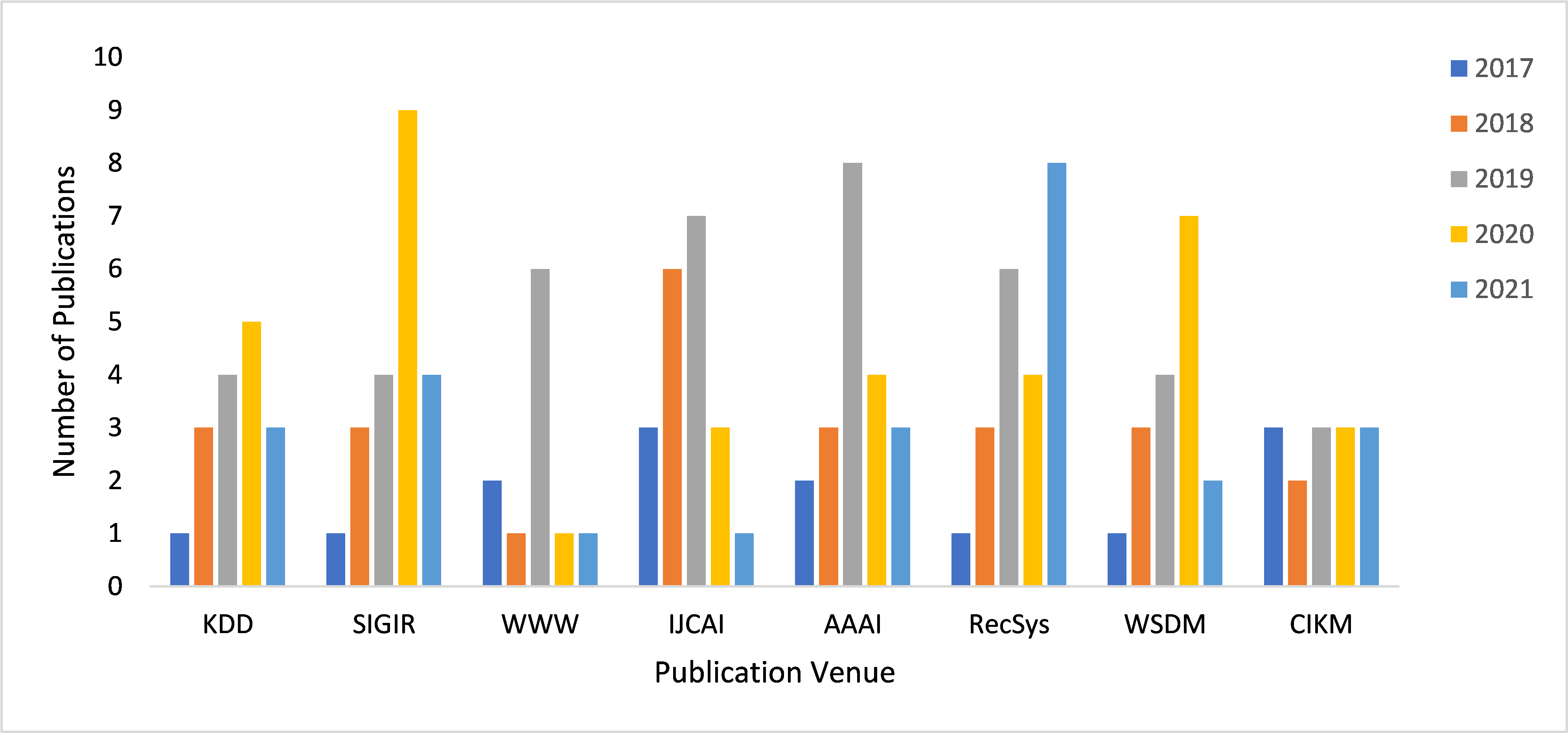}
  \caption{The Statistics of the Selected Papers.}
  \label{fig1}
\end{figure}

Specifically,
we investigate the Top-N recommendation task with implicit feedback, focusing on optimizing the benchmark recommendation algorithms commonly used in comparative experiments using hyperparameter optimization algorithms to unlock their full potential. To ensure the universality of benchmark research on recommender system algorithms, this study follows the research ideas of~\cite{sun2020we} and review those $141$ papers on top-N implicit recommendation task from eight top-tier conferences, namely RecSys, KDD, SIGIR, WWWW, IJCAI, AAAI, WSDM, and CIKM (see Figure \ref{fig1}). We set the evaluation criteria by analyzing benchmark algorithms, experimental data sets, evaluation indicators, and data preprocessing methods used in the selected $141$ papers.
Finally, in our study, the six recommendation algorithms are selected: nearest neighbor algorithm (ItemKNN)~\cite{hu2008collaborative}, pure singular value decomposition (PureSVD)~\cite{cremonesi2010performance}, matrix decomposition (MF)~\cite{rendle2012bpr}, factorization machine (FM)~\cite{rendle2010factorization}, deep learning-based (NeuMF)~\cite{he2017neural}, and graph-based (NGCF)~\cite{wang2019neural}. Besides, we explore seven widely-adopted and state-of-the-art hyperparameter optimization algorithms: random search algorithm (Random)~\cite{bergstra2012random}, simulated annealing algorithm (Anneal)~\cite{van1987simulated}, three Bayesian optimization algorithms (i.e., GPBO~\cite{snoek2012practical} based on Gaussian process, TPE~\cite{bergstra2011algorithms} based on the tree structure, and SMAC~\cite{hutter2011sequential} based on random forest), and two multi-armed bandit algorithm (i.e., Hyperband~\cite{li2017hyperband}, and BOHB~\cite{falkner2018bohb} combined with the TPE method). By conducting exhaustive experiments, our study aims to determine which hyperparameter optimization algorithm is best suited for each benchmark recommender system model by also considering different characteristics of datasets, with the goal of guiding future experiments.

The paper has two main contributions:
\begin{itemize}
\item The study contributes to fair and rigorous evaluations on recommendation systems by specifically investigating the hyperparameter optimization techniques for top-N recommendation task with implicit feedback. It provides a reliable basis for future research, enabling fair comparisons and reproducibility of results.

\item The study also facilities the model designs in the field of recommender systems by identifying the most suitable hyperparameter optimization techniques for different recommendation algorithms across different recommendation scenarios. It can serve as a reference for researchers, helping them to optimize their recommendation models and achieve better performance. 
\end{itemize}


\section{Related Work}
We mainly review two-fold of related studies: recommendation algorithms and hyperparameter optimization techniques.
\subsection{Recommendation Algorithms}
To address the information overload problem, recommender systems have been widely adopted by providing users with a list of recommended items or predictions on user preferences, thus simplifying the task of finding the most interesting items among a large set of items.
Based on the underlying ideas and the design characteristics of recommender systems, recommendation algorithms can be broadly classified into three categories \cite{sun2020we}: content-based, collaborative filtering-based, and representation learning-based recommendation algorithms.

Regarding content-based, 
Lops et al.~\cite{lops2011content} present the primary concept of content-based recommender systems, which utilize historical user-item interaction data to suggest items with similar characteristics to those previously preferred by a user. The similarity between items is determined based on their features. This type of algorithm is straightforward and aligns with human intuition. The k-nearest neighbor and item similarity-based recommendation algorithm, introduced by~\cite{hu2008collaborative}, falls under this category.
The second type of recommendation algorithms is based on collaborative filtering, which assumes that a user's behavior can be predicted based on other users who have similar behaviors. 
Various matrix decomposition techniques have been derived. For example, SVD++~\cite{koren2008factorization} is based on the basic matrix factorization model; BPRMF~\cite{rendle2012bpr} effectively incorporates user interests into model learning considerations by focusing on the order of items in the recommended lists. The factorization machine model proposed by Rendle et al.~\cite{rendle2010factorization} adds other marginal information to the MF model to help generate more accurate recommendations.
The third type is based on representation learning, which builds upon the second type and learns the embedding vector representations of users and items.
For instance, NeuMF~\cite{he2017neural} combines traditional matrix decomposition and deep neural networks, and uses user and item information to construct higher-dimensional features. NGCF~\cite{wang2019neural} introduces a graph structure to capture the collaborative signals of user-item interactions and improve the performance of the model. The Wide\&Deep model proposed by Cheng et al.~\cite{cheng2016wide} is similar to the NeuMF structure and incorporates two parts, i.e., the intersection of data features and the high-dimensional representation of features, to improve the recommendation effect of the model.

In summary, as machine learning models especially deep learning ones continue to evolve, the focus of research on recommender systems has shifted towards developing more intricate recommendation models, leading to a greater demand for effective hyperparameter optimization methods.

\subsection{Hyperparameter Optimization Techniques}

As machine learning models grow more complex, selecting the right hyperparameters is crucial. Hyperparameters, set before model training, determine model structure, training goals, and iterative algorithms. They include the number of neurons, regularization coefficients, learning rates, and so on. Optimization algorithms must adapt to different hyperparameter types. Effective tuning is vital for deep learning model performance, which has led to automatic search algorithms for hyperparameters. Automated tools have lowered the entry barrier for machine learning.

There are mainly four categories of hyperparameter optimization algorithms. The first type is sampling-based,  which aims to identify the optimal hyperparameter combination through brute-force search. Representative ones include grid search~\cite{hinton2012practical}, random search~\cite{bergstra2012random}, and Hyperband search algorithm~\cite{li2017hyperband}. Grid search divides the hyperparameter search space equally and combines the samples systematically to generate sets of hyperparameter candidates. Random search randomly selects sample points in the search space to create hyperparameter candidate sets. The Hyperband search algorithm enhances the search efficiency by utilizing the random search approach to allocate computing resources and terminating the evaluation of poorly performing hyperparameter combinations ahead of time.
Greedy-based algorithms make up the second type, like the mountain climbing and simulated annealing algorithms~\cite{van1987simulated}. Simulated annealing algorithm is bulit on the hill climbing algorithm's complete greediness by adding disturbances, thereby preventing the algorithm from remaining in a local optimal solution.

The third type is the model-based, which involves replacing the original model with a simpler surrogate model like a deep neural network that is easy to evaluate. This surrogate model takes information from the candidate set and constructs an acquisition function that balances between exploitation and exploration. The hyperparameter combination with the highest potential to achieve the best performance is chosen based on the candidate set's performance on the acquisition function, and then fully evaluated on the original complex model. Different model-based hyperparameter search algorithms utilize different surrogate models and acquisition functions. Notable examples include the Bayesian optimization GPBO algorithm that uses a Gaussian process~\cite{snoek2012practical,hutter2010time,wang2016bayesian} as its surrogate model, as well as the SMAC algorithm that employs a random forest~\cite{hutter2011sequential}, and the TPE algorithm that relies on the tree structure Parzen estimation method~\cite{bergstra2011algorithms} .

Finally, the fourth type refers to hyper-gradient-based~\cite{franceschi2017forward,pedregosa2016hyperparameter,shi2021improved}, which stands out from the other three categories because it constructs the gradient for the hyperparameters using the outer layer structural information of the optimization problem. To discover the optimal combination, this algorithm utilizes conventional gradient-based optimization algorithms to iteratively update the hyperparameters. However, this type of algorithms uses the backward propagation method to calculate hypergradients, which are resource-intensive and time-consuming. Various hyperparameter optimization algorithms that rely on hypergradients vary in terms of their precision and time-space complexity in representing the hypergradient. For example, \cite{bengio2000gradient} initially incorporated backpropagation into hyperparameter optimization, while \cite{shi2021improved} reformulated the original problem as a constrained optimization problem, effectively solving the dual-layer optimization issue.
Furthermore, a prevalent method for emulating manual hyperparameter tuning is to monitor the learning curve of an iterative algorithm to assess the effectiveness of the current hyperparameter combination. If the performance does not show any significant improvement, the training process is terminated early. The concept of early termination training was initially introduced in the hyperparameter search process by~\cite{domhan2015speeding}. In contrast,~\cite{klein2017learning} utilized Bayesian networks to forecast the fitting performance of hyperparameter combinations.

To conclude, our research aims to investigate suitable hyperparameter optimization methods for various recommendation algorithms in diverse recommendation scenarios.

\section{Experimental Design}
\subsection{Training Procedure}
\begin{figure}[htbp]
  \centering
  \includegraphics[width=\linewidth]{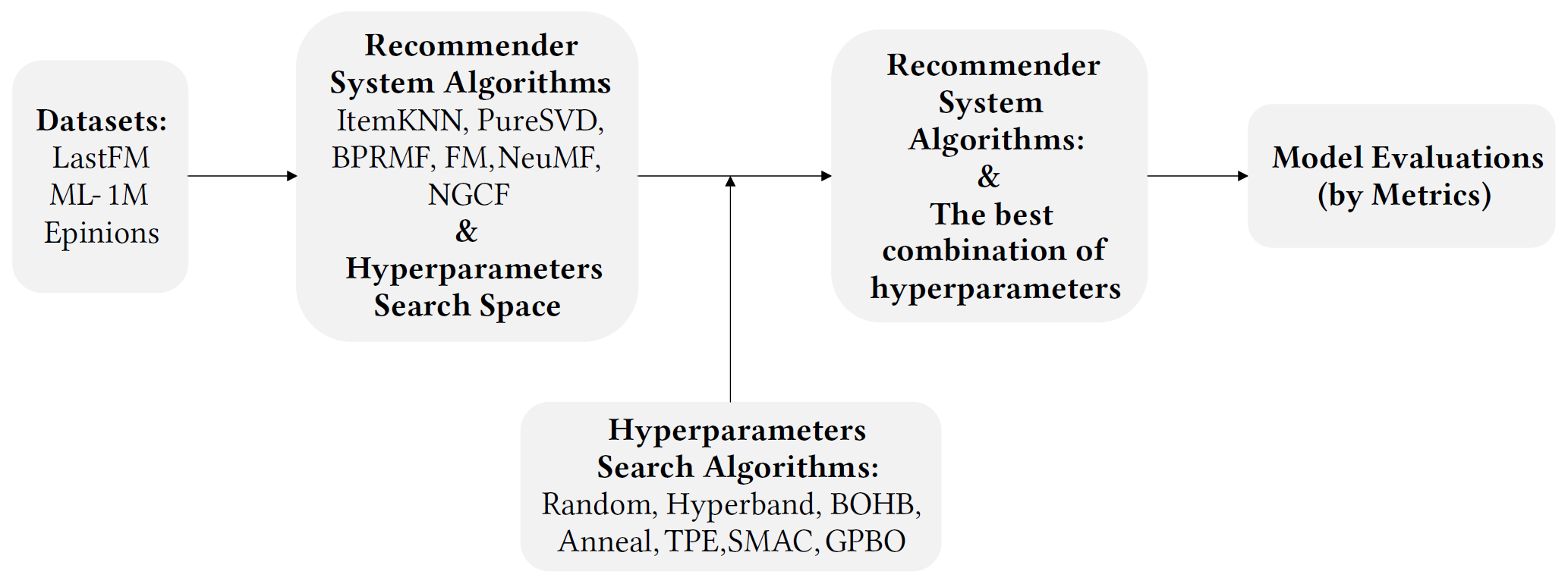}
  \caption{The Training Procedure of Different Models.}
  \label{fig5}
\end{figure}

In our study, the training procedure is depicted in Figure \ref{fig5}. We aim to optimize a selected benchmark recommendation model (e.g., NGCF) defined by a set of hyperparameters and an specifically designed algorithm on every training set of selected datasets. The trained model is then evaluated using the corresponding validation set measured by an evaluation metric, which is a data tuple <hyperparameter combination, evaluation index>. This process is repeated for multiple hyperparameter combinations, resulting in a set of data tuples. Finally, the hyperparameter combination that yields the best performance is selected for complete model training, and the performance on the corresponding test set is finally evaluated using this best model.

As can be seen in Figure \ref{fig5}, we first select representative datasets and split every dataset into training set, validation set and test set, respectively. We also identify six representative and widely-adopted recommendation algorithms for top-N recommendation task with implicit feedback. We then tune these methods on each dataset using seven popular hyperparameter optimization techniques.
Next, we will elaborate each component in Figure \ref{fig5} in great details.

\subsection{Dataset and Preprocessing}
Recommender system algorithms have been widely employed across various platforms such as movies, music, and other domains. In order to meet the criterion of high domain coverage and data volume of the experimental data, following \cite{sun2020we}, our study selects three experimental datasets, music dataset LastFM, movies dataset MovieLens-1M, and e-commerce review dataset Epinions. Table \ref{tab1} presents the statistical information of the three selected experimental datasets, and also delves into the distribution of user interaction records (\emph{data density}) within the entire datasets, thereby facilitating subsequent experiment analyses.

\begin{table}[htbp]
  \caption{Statistical Information of Experimental Datasets.}
  \label{tab:freq}
  \begin{tabular}{lccc}
    \toprule
    Datasets &LastFM &ML-1M &Epinions\\
    \midrule
    \#users & 1,892 & 6,038 & 22,164\\
    \#items & 17,632 & 3,533 & 296,277\\
    \#interactions & 92,834 & 575,281 & 922,267\\
    sparsity & 2.783e-03 & 2.697e-02 & 1.404e-04\\
     \#interaction =1 & 23.08\% & 0.16\% & 40.35\% \\
    \#interaction <=5 & 9.08\% & 0.49\% & 12.57\% \\
    \#interaction >5 & 67.85\% & 99.35\% & 47.08\% \\
  \bottomrule
\end{tabular}
\label{tab1}
\end{table}

Our study targets at the Top-N implicit recommendation task, which assigns $1$ to a user-item interaction if the interaction exists between the user and the item, and $0$ otherwise. 
Among the three experimental datasets selected, interactions on LastFM dataset are inherently implicit, while those on ML-1M and Epinions datasets are explicit scores. We then apply a preprocessing method whereby scores greater than $4$ are converted to 1, while those less than or equal to $4$ are converted to $0$.

Besides, during model training, we employ the Uniform Sampler method~\cite{he2017neural} which extracts a number of items with equal probability from the set of items that have not been interacted with by every user as negative samples. The number of negative samples is a critical one-dimensional hyperparameter that affects the performance of the model and must be tuned using the hyperparameter search algorithm chosen in this study.  


\subsection{Data Splitting Method}
There are mainly two types of data splitting methods: split-by-ratio and leave-one-out. In particular, split-by-ratio means that a proportion $\rho$ (e.g., $\rho=80\%$) of the dataset (i.e., user-item interaction records) is treated as training set, and the rest ($1-\rho=20\%$) as test set. While, leave-one-out refers to that for each user, only one record is kept for test and the remaining are for training. Besides, some papers directly split training and test sets by a fixed timestamp, that is, the data before the fixed timestamp is used as training set, and the rest as test set.

In our study, we follow the majority practice and select both randomand time-aware split-by-ratio at global-level with $\rho=80\%$ as our data splitting method. Meanwhile, to speed up the test process, we randomly sample negative items for each user to ensure her test candidates to be 1,000, and rank all test items among the 1,000 items.

Besides the split for training and test sets in cross-validation, an extra validation set should be retained to help tune the hyperparameters. We adopt the nested validation strategy, that is, in each fold of cross-validation, we further select from the training set 10\% of records as the validation set to tune hyper-parameters. Once the optimal settings for hyper-parameters are decided, we feed the whole training set (including the validation set) to train the final model and then report the performance on the test set.

\subsection{Recommendation Algorithms}
Following \cite{sun2020we} and after examining the $141$ papers, we select six representative and popular algorithms from the three categories for implicit top-N recommendation task:
\subsubsection{Content-based Recommendation Algorithms}
\begin{itemize}
    \item[$\bullet$]\textbf{ItemKNN}~\cite{hu2008collaborative} is an K-nearest neighborhood based method which recommends items based on item similarity (i.e., cosine similarity).
\end{itemize}
\subsubsection{Collaborative Filtering-based Recommendation Algorithms}
\begin{itemize}
\item[$\bullet$]\textbf{PureSVD}~\cite{cremonesi2010performance} is a matrix factorization-based algorithm which conducts Singular Value Decomposition (SVD) on user-item interaction matrix.
\item[$\bullet$]\textbf{BPRMF}~\cite{rendle2012bpr} is a recommendation algorithm based on Bayesian personalized ranking matrix factorization (BPRMF), and its goal is to maximize the posterior probability, making visited items superior to unvisited items.
\item[$\bullet$]\textbf{FM}~\cite{rendle2010factorization} combines the advantages of Support Vector Machines (SVM) with factorization models, and considers both explicit features and implicit user-item interaction matrix.
\end{itemize}
\subsubsection{Representation Learning-based Recommendation Algorithms}
\begin{itemize}
    \item[$\bullet$]\textbf{NeuMF}~\cite{he2017neural} uses a multi-layer perceptron architecture to learn the latent representations of users and items and thus further improves recommendation performance.
    \item[$\bullet$]\textbf{NGCF}~\cite{wang2019neural} exploits the user-item graph structure and leverages graph neural network to better learn the high-order connectivity relationship between users and items.
\end{itemize}

\subsection{Evaluation Metrics}
Following \cite{sun2020we}, we select the two representative and widely adopted evaluation metrics in our study, that is, NDCG (Normalized Discounted Cumulative Gain) and HR (Hit Ratio).
In particular, they both intuitively measure whether a test item is present in the top-N recommendation list and higher values indicate better model performance, whilst NDCG also accounts for the ranking positions of test items and normalizes the evaluation scores across different users.
Besides, we consider $N=5$ and $N=10$ in the following experiments.
Furthermore, in the model-based hyperparameter optimization algorithms, we choose NDCG as the criterion to guide the selection of model hyperparameters.

\subsection{Hyperparameter Search Algorithms}
We consider seven representative and stat-of-the-art hyperparameter optimization techniques in our study to tune hyperparameters for different recommendation algorithms.
\subsubsection{Sampling-based Hyperparameter Search Algorithms}
\begin{itemize}   \item[$\bullet$]\textbf{Random}~\cite{bergstra2012random} randomly selects hyperparameters from predefined search spaces and then evaluates their performance.

\item[$\bullet$]\textbf{Hyperband}~\cite{li2017hyperband} is a multi-armed bandit-based approach for hyperparameter optimization which speeds up random search through adaptive resource allocation and early-stopping.

\item[$\bullet$]\textbf{BOHB}~\cite{falkner2018bohb} combines the strengths of Bayesian optimization and Hyperband. It uses Bayesian optimization to select promising configurations and Hyperband to speed up the evaluation process through adaptive resource allocation and early-stopping.
\end{itemize}

\subsubsection{Greedy-based Hyperparameter Search Algorithm}
\begin{itemize}
    \item[$\bullet$]\textbf{Anneal}~\cite{van1987simulated} is a well-studied local search method which provides a mechanism to escape local optima by allowing hill-climbing moves (i.e., moves which worsen the objective function value) in hopes of finding a global optimum.
\end{itemize}

\subsubsection{Model-based Hyperparameter Search Algorithms}
\begin{itemize}
    \item[$\bullet$]\textbf{GPBO}~\cite{snoek2012practical} is a Bayesian optimization algorithm, which uses a Gaussian process as a surrogate model and can only accept continuous hyperparameter search spaces.
    \item[$\bullet$]\textbf{SMAC}~\cite{hutter2011sequential} is a Bayesian optimization algorithm which uses random forest as a surrogate model and resolves hyperparameter search limitations for categorical variables.
    \item[$\bullet$]\textbf{TPE}~\cite{bergstra2011algorithms} is a Bayesian optimization algorithm that uses a Gaussian mixture model as a surrogate model. It uses a tree-structured Parzen estimator to model the objective function, and estimates a probability distribution of hyperparameters from which to choose the next set of hyperparameters to optimize.
\end{itemize}

\subsection{Hyperparameter Settings}
It is essential to set the search space for hyperparameters before training a recommendation algorithm. Different algorithms contain unique hyperparameters, which require manual specification based on the algorithm's structural characteristics and the research goal. Table \ref{tab3} summarizes the hyperparameters and search space configurations used in this study. Four hyperparameters are set for both BPRMF and BPRFM models, whereas NeuMF and NGCF involve seven hyperparameters. Additionally, the data types of hyperparameters vary, including integer, floating-point, and category types.

\begin{table}[ht] 
  \caption{Hyperparameter Configurations for Recommendation Models.}
    \begin{tabular}{|c|l|l|l|l|}
    \hline
    \textbf{Recommender Systems} & \textbf{Hyperparameters} & \textbf{Type} & \textbf{Search Range} & \textbf{Explanation} \\ \hline
    ItemKNN	& -maxk	& Int & [1, 100] & Number of item neighbors\\ \hline
    PureSVD	& -factors & Int & [1, 100] & Latent factor dimension\\ \hline
    \multirow{4}{*}{BPRMF} & -num\_ng & Int & [1, 10] & Number of negative samples \\  
                           & -factors & Int & [1, 100] & Latent factor dimension \\  
                           & -lr & Float & [1e-4, 1e-2] & Learning rate \\   
                           & -reg\_2 & Float & [1e-4, 1e-2] & Regularization coefficient \\ \hline
    \multirow{4}{*}{FM} & -num\_ng & Int & [1, 10] & Number of negative samples \\  
                        & -factors & Int & [1, 100] & Latent factor dimension \\  
                        & -lr & Float & [1e-4, 1e-2] & Learning rate \\   
                        & -reg\_2 & Float & [1e-4, 1e-2] & Regularization coefficient \\ \hline
    \multirow{7}{*}{NeuMF} & -num\_ng & Int & [1, 10] & Number of negative samples \\  
                           & -factors & Int & [1, 100] & Latent factor dimension \\  
                           & -num\_layers & Int & [1, 3] & Number of layers \\  
                           & -dropout & Float & [0, 1] & Dropout rate \\   
                           & -lr & Float & [1e-4, 1e-2] & Learning rate \\   
                           & -reg\_2 & Float & [1e-4, 1e-2] & Regularization coefficient \\
                          & -batch\_size & Int & \{64, 128, 256, 512\} &Batch size\\ \hline
    \multirow{7}{*}{NGCF} & -num\_ng & Int & [1, 10] & Number of negative samples \\  
                          & -factors & Int & [1, 100] & Latent factor dimension \\   
                          & -lr & Float & [1e-4, 1e-2] & Learning rate \\   
                          & -reg\_2 & Float & [1e-4, 1e-2] & Regularization coefficient \\
                          & -node\_dropout & Float & [1e-4, 1e-2] & Node dropout rate \\
                          & -mess\_dropout & Float & [1e-4, 1e-2] & Message dropout rate \\
                          & -batch\_size & Int & \{64, 128, 256, 512\} & Batch size\\ \hline
    \end{tabular}
    \label{tab3}
\end{table}

We then employ the aforementioned seven hyperparameter optimization techniques to tune hyperparameters for each model.
Specifically, we sample $20$ sets of hyperparameter combinations according to its own logic for each hyperparameter optimization method, and then train each recommendation model for 30 epochs on each data set. The Hyperband and BOHB algorithms also require pre-setting of computing resources, where here the epoch of model training was chosen as the computing resource to be allocated, with a minimum of $5$ epochs and a maximum of $30$ epochs. Stepwise training is performed.

Finally, to eliminate the varied affect caused by different objective functions across different recommendation models \cite{sun2020we}, we choose the same objective function for different recommendation models, i.e., a pair-wise BPR loss.

\section{Experimental Evaluations}
\subsection{Recommendation Performance}
\begin{figure}[htbp]
  \centering
  \includegraphics[width=0.95\linewidth]{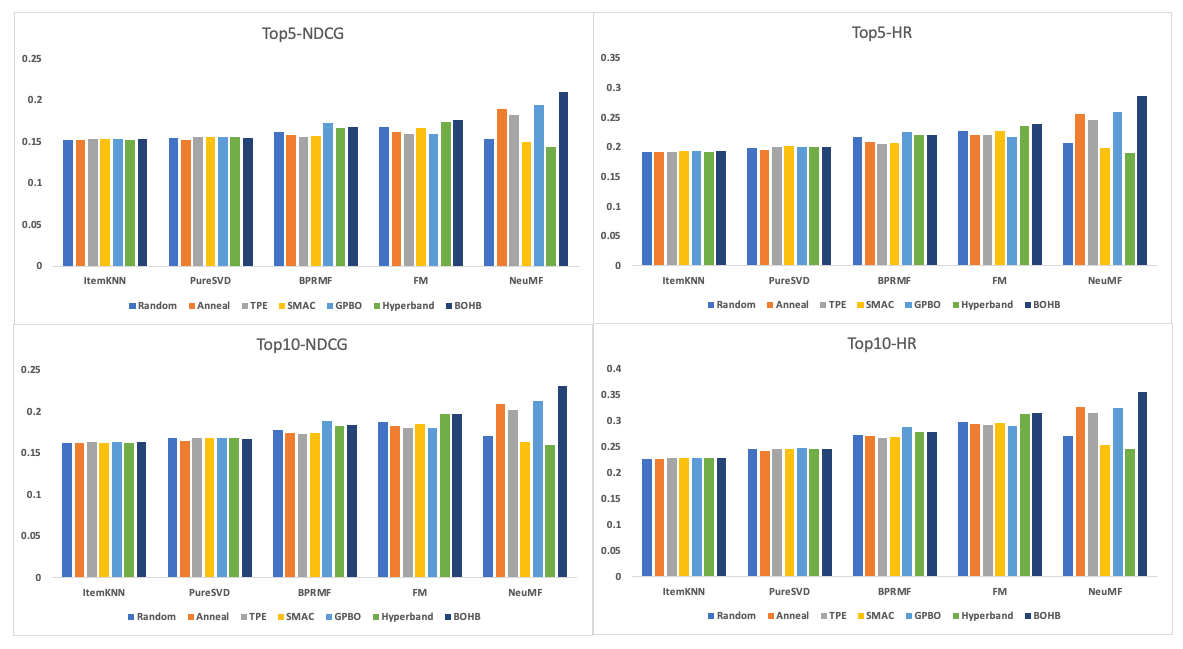}
  \caption{Experimental Results on Epinions.}
  \label{fig7}
\end{figure}

Table 
\ref{tb:lastfm} summarizes the experimental results in terms of NDCG and HR on LastFM. Table \ref{tb:ml1m} presents experimental results on ML-1M, whilst those on Epinions are presented in Figure \ref{fig7}. All the results in tables are presented in the form of ``$\text{mean}\pm\text{deviation}$" based on three rounds of experiments.
We have the following observations:

(1) on LastFM, the ascending order of recommendation accuracy is as follows: FM, NeuMF, BPRMF, PureSVD, ItemKNN, and NGCF, whilst on ML-1M and Epinions, the corresponding order is ItemKNN, PureSVD, BPRMF, FM, NGCF, and NeuMF. On LastFM, the simpler recommendation algorithms, PureSVD and ItemKNN, outperform the more complex algorithms like NeuMF and BPRMF after thorough calibration of hyperparameters. It is worth mentioning that the size of LastFM is significantly smaller than that of ML-1M and Epinions. Hence, it can be inferred that for smaller experimental datasets, utilizing simpler recommendation algorithms like ItemKNN and PureSVD, in combination with hyperparameter optimization techniques for thorough tuning, can yield better outcomes. However, for larger experimental datasets, a more sophisticated recommender system model that automatically captures more information can still produce superior recommendation results.
Besides,
the level of data sparsity and data density across the three datasets ranks from low to high as Epinions, LastFM, and ML-1M, which does not align with the ranking of their recommendation performance. That is, for example, the performance of these algorithms on ML-1M (with higher data density) is worse that on LastFM (with lower data density). 

\begin{table}[htbp]
  \centering
  \caption{Experimental Results on LastFM. Best Result is Highlighted in Boldface and the Worst One is Underlined.}
    \begin{tabular}{c|c|llllll}
    \toprule
 Metric&
    \multicolumn{1}{c|}{Algorithms} & \multicolumn{1}{c}{ItemKNN} & \multicolumn{1}{c}{PureSVD} & \multicolumn{1}{c}{BPRMF} & \multicolumn{1}{c}{FM} & \multicolumn{1}{c}{NeuMF} & \multicolumn{1}{c}{NGCF}  \\
    \midrule
    \multirow{8}{*}{NDCG@5}&
    Random & 0.819$\pm$0.002 & 0.806$\pm$0.002 & 0.744$\pm$0.011 & 0.589$\pm$0.002 & 0.730$\pm$0.019 & 0.823$\pm$0.012 \\
    &Anneal & 0.817$\pm$0.001 & \textbf{0.810$\pm$0.000} & 0.776$\pm$0.013 & 0.589$\pm$0.002 & 0.712$\pm$0.039 & 0.820$\pm$0.004 \\
    &TPE & 0.819$\pm$0.001 & 0.808$\pm$0.001 & 0.752$\pm$0.022 & 0.587$\pm$0.001 & 0.711$\pm$0.017 & 0.817$\pm$0.007 \\
    &SMAC & \textbf{0.820$\pm$0.002} & \underline{0.805$\pm$0.003} & 0.768$\pm$0.031 & 0.592$\pm$0.013 & 0.729$\pm$0.009 & 0.810$\pm$0.013 \\
    &GPBO & 0.818$\pm$0.001 & 0.808$\pm$0.001 & \underline{0.733$\pm$0.092} & \underline{0.585$\pm$0.006} & \underline{0.676$\pm$0.090} & \underline{0.756$\pm$0.062} \\
    &Hyperband & 0.817$\pm$0.003 & 0.809$\pm$0.002 & 0.781$\pm$0.004 & \textbf{0.653$\pm$0.081} & 0.731$\pm$0.049 & \textbf{0.834$\pm$0.004} \\
    &BOHB & \underline{0.815$\pm$0.001} & 0.808$\pm$0.004 & \textbf{0.791$\pm$0.006} & 0.640$\pm$0.073 & \textbf{0.778$\pm$0.025} & 0.833$\pm$0.009 \\

  \midrule
  \multirow{8}{*}{HR@5}
    &Random & 0.929$\pm$0.002 & \textbf{0.918$\pm$0.003} & 0.875$\pm$0.013 & 0.748$\pm$0.003 & 0.874$\pm$0.014 & 0.942$\pm$0.005 \\
    &Anneal & 0.928$\pm$0.001 & 0.810$\pm$0.000 & 0.903$\pm$0.011 & 0.750$\pm$0.001 & 0.851$\pm$0.041 & 0.942$\pm$0.003 \\
    &TPE & 0.930$\pm$0.001 & 0.914$\pm$0.003 & 0.884$\pm$0.020 & \underline{0.747$\pm$0.002} & 0.864$\pm$0.013 & 0.938$\pm$0.004 \\
    &SMAC & \textbf{0.933$\pm$0.003} & 0.917$\pm$0.002 & 0.897$\pm$0.024 & 0.755$\pm$0.009 & 0.872$\pm$0.006 & 0.939$\pm$0.003 \\
    &GPBO & \underline{0.927$\pm$0.000} & 0.913$\pm$0.001 & \underline{0.869$\pm$0.075} & 0.749$\pm$0.006 & \underline{0.818$\pm$0.097} & \underline{0.900$\pm$0.034} \\
    &Hyperband & 0.932$\pm$0.002 & 0.917$\pm$0.002 & 0.906$\pm$0.000 & \textbf{0.800$\pm$0.068} & 0.876$\pm$0.033 & 0.946$\pm$0.003 \\
    &BOHB & 0.927$\pm$0.001 & \underline{0.912$\pm$0.004} & \textbf{0.916$\pm$0.002} & 0.791$\pm$0.060 & \textbf{0.911$\pm$0.013} & \textbf{0.948$\pm$0.004} \\

    \midrule
    \multirow{8}{*}{NDCG@10}
    &Random & 0.807$\pm$0.002 & \underline{0.796$\pm$0.001} & 0.742$\pm$0.009 & 0.607$\pm$0.002 & 0.729$\pm$0.019 & 0.812$\pm$0.011 \\
    &Anneal & 0.806$\pm$0.001 & 0.800$\pm$0.000 & 0.769$\pm$0.011 & 0.606$\pm$0.001 & 0.713$\pm$0.032 & 0.807$\pm$0.003 \\
    &TPE & 0.807$\pm$0.001 & 0.799$\pm$0.001 & 0.747$\pm$0.019 & 0.605$\pm$0.001 & 0.714$\pm$0.016 & 0.805$\pm$0.007 \\
    &SMAC & \textbf{0.810$\pm$0.001} & 0.796$\pm$0.002 & 0.761$\pm$0.026 & 0.608$\pm$0.010 & 0.725$\pm$0.006 & 0.801$\pm$0.010 \\
    &GPBO & 0.807$\pm$0.001 & 0.799$\pm$0.001 & \underline{0.731$\pm$0.081} & \underline{0.603$\pm$0.004} & \underline{0.679$\pm$0.081} & \underline{0.758$\pm$0.054} \\
    &Hyperband & 0.805$\pm$0.002 & \textbf{0.800$\pm$0.002} & 0.771$\pm$0.003 & \textbf{0.661$\pm$0.069} & 0.729$\pm$0.044 & \textbf{0.822$\pm$0.003} \\
    &BOHB & \underline{0.804$\pm$0.001} & 0.799$\pm$0.003 & \textbf{0.783$\pm$0.004} & 0.651$\pm$0.065 & \textbf{0.771$\pm$0.023} & 0.821$\pm$0.009 \\

    \midrule
    \multirow{8}{*}{HR@10}
    &Random & 0.959$\pm$0.002 & 0.948$\pm$0.002 & 0.932$\pm$0.006 & 0.840$\pm$0.001 & 0.932$\pm$0.014 & 0.969$\pm$0.003 \\
    &Anneal & \underline{0.957$\pm$0.001} & \textbf{0.949$\pm$0.001} & 0.946$\pm$0.006 & 0.840$\pm$0.002 & 0.913$\pm$0.024 & 0.967$\pm$0.001 \\
    &TPE & 0.960$\pm$0.002 & \underline{0.946$\pm$0.001} & 0.933$\pm$0.013 & 0.839$\pm$0.004 & 0.927$\pm$0.011 & 0.965$\pm$0.002 \\
    &SMAC & \textbf{0.961$\pm$0.001} & 0.948$\pm$0.002 & 0.940$\pm$0.014 & 0.845$\pm$0.003 & 0.923$\pm$0.001 & 0.966$\pm$0.001 \\
    &GPBO & 0.958$\pm$0.001 & 0.947$\pm$0.002 & \underline{0.920$\pm$0.052} & \underline{0.838$\pm$0.002} & \underline{0.882$\pm$0.077} & \underline{0.951$\pm$0.016} \\
    &Hyperband & 0.958$\pm$0.001 & 0.947$\pm$0.001 & 0.943$\pm$0.003 & \textbf{0.873$\pm$0.045} & 0.928$\pm$0.021 & 0.971$\pm$0.001 \\
    &BOHB & 0.959$\pm$0.002 & 0.946$\pm$0.002 & \textbf{0.953$\pm$0.001} & 0.870$\pm$0.043 & \textbf{0.949$\pm$0.006} & \textbf{0.972$\pm$0.003} \\
    \bottomrule
    \end{tabular}%
 \label{tb:lastfm}
\end{table}%

\begin{table*}[htbp]
  \centering
  \caption{Experimental Results on ML-1M. Best Result is Highlighted in Boldface and the Worst One is Underlined.}
    \begin{tabular}{c|c|llllll}
    \toprule
    Metric& \multicolumn{1}{c|}{Algorithms} & \multicolumn{1}{c}{ItemKNN} & \multicolumn{1}{c}{PureSVD} & \multicolumn{1}{c}{BPRMF} & \multicolumn{1}{c}{FM} & \multicolumn{1}{c}{NeuMF} & \multicolumn{1}{c}{NGCF}  \\
    \midrule
    \multirow{8}{*}{NDCG@5}
    &Random & 0.394$\pm$0.005 & \textbf{0.424$\pm$0.006} & 0.480$\pm$0.004 & 0.642$\pm$0.015 & 0.628$\pm$0.026 & 0.447$\pm$0.091 \\
    &Anneal & 0.404$\pm$0.003 & 0.417$\pm$0.006 & \textbf{0.493$\pm$0.002} & \underline{0.637$\pm$0.007} & \underline{0.528$\pm$0.159} & \textbf{0.589$\pm$0.006} \\
    &TPE & \underline{0.394$\pm$0.002} & 0.424$\pm$0.004 & 0.480$\pm$0.007 & 0.642$\pm$0.012 & 0.648$\pm$0.005 & 0.574$\pm$0.025 \\
    &SMAC & 0.402$\pm$0.004 & \underline{0.415$\pm$0.005} & 0.487$\pm$0.002 & \textbf{0.654$\pm$0.012} & 0.536$\pm$0.116 & \underline{0.378$\pm$0.032} \\
    &GPBO & 0.397$\pm$0.005 & 0.421$\pm$0.008 & 0.493$\pm$0.010 & 0.645$\pm$0.011 & 0.635$\pm$0.012 & 0.586$\pm$0.048 \\
    &Hyperband & \textbf{0.405$\pm$0.003} & 0.422$\pm$0.001 & \underline{0.473$\pm$0.008} & 0.642$\pm$0.008 & 0.645$\pm$0.015 & 0.582$\pm$0.013 \\
    &BOHB & 0.398$\pm$0.004 & 0.416$\pm$0.005 & 0.475$\pm$0.018 & 0.640$\pm$0.006 & \textbf{0.649$\pm$0.003} & 0.573$\pm$0.039 \\

    \midrule
    \multirow{8}{*}{HR@5}
    &Random & 0.528$\pm$0.003 & 0.561$\pm$0.006 & 0.631$\pm$0.007 & 0.776$\pm$0.013 & 0.763$\pm$0.021 & 0.589$\pm$0.128 \\
    &Anneal & 0.536$\pm$0.004 & 0.560$\pm$0.008 & 0.641$\pm$0.002 & \underline{0.774$\pm$0.005} &\underline{0.668$\pm$0.160} & \textbf{0.759$\pm$0.001} \\
    &TPE & \underline{0.524$\pm$0.003} & \textbf{0.565$\pm$0.006} & 0.630$\pm$0.008 & 0.782$\pm$0.013 & \textbf{0.789$\pm$0.007} & 0.737$\pm$0.028 \\
    &SMAC & \textbf{0.537$\pm$0.005} & 0.549$\pm$0.005 & 0.635$\pm$0.004 & 0.788$\pm$0.015 & 0.672$\pm$0.125 & \underline{0.521$\pm$0.049} \\
    &GPBO & 0.529$\pm$0.005 & 0.553$\pm$0.009 & \textbf{0.648$\pm$0.012} & 0.782$\pm$0.013 & 0.768$\pm$0.019 & 0.746$\pm$0.041 \\
    &Hyperband & 0.534$\pm$0.006 & 0.559$\pm$0.010 & \underline{0.612$\pm$0.020} & 0.778$\pm$0.012 & 0.782$\pm$0.020 & 0.751$\pm$0.008 \\
    &BOHB & 0.532$\pm$0.005 & \underline{0.546$\pm$0.006} & 0.625$\pm$0.024 & 0.779$\pm$0.010 & 0.788$\pm$0.003 & 0.733$\pm$0.035 \\

    \midrule
    \multirow{8}{*}{NDCG@10}
    &Random & 0.430$\pm$0.004 & \textbf{0.457$\pm$0.005} & 0.511$\pm$0.006 & 0.660$\pm$0.012 & 0.647$\pm$0.022 & 0.418$\pm$0.098 \\
    &Anneal & 0.439$\pm$0.002 & \underline{0.450$\pm$0.005} & \textbf{0.517$\pm$0.003} & \underline{0.655$\pm$0.004} & \underline{0.553$\pm$0.147} & \textbf{0.616$\pm$0.006} \\
    &TPE & \underline{0.432$\pm$0.001} & 0.457$\pm$0.004 & 0.507$\pm$0.011 & 0.661$\pm$0.011 & 0.664$\pm$0.004 & 0.605$\pm$0.019 \\
    &SMAC & 0.437$\pm$0.002 & 0.450$\pm$0.004 & 0.512$\pm$0.001 & \textbf{0.668$\pm$0.010} & 0.558$\pm$0.108 & \underline{0.414$\pm$0.041} \\
    &GPBO & 0.433$\pm$0.003 & 0.455$\pm$0.006 & 0.516$\pm$0.010 & 0.663$\pm$0.008 & 0.650$\pm$0.014 & 0.612$\pm$0.042 \\
    &Hyperband & \textbf{0.440$\pm$0.002} & 0.454$\pm$0.003 & \underline{0.502$\pm$0.008} & 0.660$\pm$0.008 & 0.660$\pm$0.014 & 0.609$\pm$0.013 \\
    &BOHB & 0.434$\pm$0.002 & 0.452$\pm$0.005 & 0.502$\pm$0.017 & 0.658$\pm$0.004 & \textbf{0.665$\pm$0.003} & 0.603$\pm$0.032 \\

    \midrule
    \multirow{8}{*}{HR@10}
    &Random & \underline{0.648$\pm$0.005} & 0.674$\pm$0.001 & \textbf{0.748$\pm$0.012} & 0.857$\pm$0.008 & 0.845$\pm$0.015 & 0.706$\pm$0.144 \\
    &Anneal & 0.654$\pm$0.002 & 0.676$\pm$0.009 & 0.742$\pm$0.006 & \underline{0.853$\pm$0.002} & 0.764$\pm$0.131 & \textbf{0.851$\pm$0.002} \\
    &TPE & 0.648$\pm$0.008 & \textbf{0.679$\pm$0.007} & 0.736$\pm$0.018 & \textbf{0.864$\pm$0.010} & \textbf{0.865$\pm$0.006} & 0.845$\pm$0.013 \\
    &SMAC & 0.651$\pm$0.002 & 0.669$\pm$0.004 & 0.732$\pm$0.011 & 0.861$\pm$0.009 & \underline{0.761$\pm$0.109} & \underline{0.639$\pm$0.077} \\
    &GPBO & 0.649$\pm$0.006 & 0.671$\pm$0.005 & 0.747$\pm$0.013 & 0.861$\pm$0.010 & 0.846$\pm$0.015 & 0.840$\pm$0.029 \\
    &Hyperband & \textbf{0.654$\pm$0.003} & 0.669$\pm$0.004 & \underline{0.713$\pm$0.034} & 0.858$\pm$0.011 & 0.857$\pm$0.014 & 0.842$\pm$0.007 \\
    &BOHB & \textbf{0.654$\pm$0.003} & \underline{0.668$\pm$0.007} & 0.728$\pm$0.023 & 0.862$\pm$0.006 & 0.864$\pm$0.006 & 0.835$\pm$0.019 \\
    \bottomrule
    \end{tabular}%
 \label{tb:ml1m}
\end{table*}%

(2) with regards to the use of hyperparameter optimization algorithms for different recommendation algorithms, the Anneal and TPE algorithms have demonstrated better performance when used to tune ItemKNN, PureSVD, BPRMF, and FM algorithms. However, for more intricate algorithms like NeuMF and NGCF, it is advisable to use Hyperband and BOHB search algorithms for tuning. This is because NeuMF and NGCF have more complex model structures, longer model fitting times, a larger number of hyperparameters to optimize, and more complex data types. In terms of experimental efficiency, the Hyperband and BOHB algorithms exhibit better results and are more suitable. However, BOHB's time efficiency is lower compared to Hyperband due to its utilization of a model-based algorithm. If high search efficiency is desired, Hyperband is recommended, and if high performance is desired, BOHB is recommended. Random search tuning is not advisable due to its less reliable performance.

(3) the GPBO algorithm can only handle continuous data types, so it needs to serialize and sample integer-type hyperparameters before discretizing them. However, this approach adds an approximation step, which leads to deviations and affects the final outcome. Moreover, the GPBO algorithm cannot tune categorical hyperparameters. The experimental results show that some algorithms like ItemKNN, PureSVD, BPRMF, and FM have discrete hyperparameters, while others (e.g., NeuMF and NGCF) have both discrete and categorical hyperparameters. Therefore, the GPBO algorithm does not work well in tuning these models. Based on this analysis, it is not recommended to use the GPBO algorithm to tune these six benchmark models.

To summarize, for ItemKNN, PureSVD, BPRMF, and FM algorithms, Anneal and TPE should be used for tuning, whereas Hyperband and BOHB are recommended for more complex models like NeuMF and NGCF. Random search tuning is not recommended as it may lead to less robust results, and GPBO tuning should also be avoided as it may not scale well with hyperparameter types.

\subsection{Analysis of Hyperparameter Search Process}
In this section, we strive to explore the hyperparameter search process of different hyperparameter optimization techniques on recommendation scenario. Particularly, due to space limitation, we take BPRMF model on LastFM as example in this paper. Similar results can be obtained for other recommendation algorithms on other scenarios. 
Details of the optimal hyperparameters obtained by different hyperparameter optimiztion algorithms are presented in Table \ref{tab12}. It can be seen that the range of hyperparameter values in each dimension is clearly evident. When dealing with a combinatorial optimization problem like this, attempting manual tuning would not be efficient or accurate enough to achieve this level.


\begin{table*}[htbp]
  \centering
  \caption{Best Hyperparameters of the BPRFM Model under Each Optimization Algorithm on LastFM.}
    \begin{tabular}{c|llll}
    \toprule
    \multicolumn{1}{c}{} & \multicolumn{1}{c}{num\_ng} & \multicolumn{1}{c}{factors} & \multicolumn{1}{c}{lr} & \multicolumn{1}{c}{reg\_2} \\
    \midrule
    Anneal & 6 & 77 & 0.0084 & 0.0019  \\
    Random & 8 & 96 & 0.0053 & 0.0083  \\
    TPE & 5 & 63 & 0.0086 & 0.0004  \\
    SMAC & 8 & 100 & 0.0083 & 0.0055  \\
    GPBO & 10 & 88 & 0.0100 & 0.0001 \\
    Hyperband & 10 & 98 & 0.0036 & 0.0008 \\
    BOHB & 10 & 40 & 0.0070 & 0.0002 \\
    \bottomrule
    \end{tabular}%
 \label{tab12}
\end{table*}%

\begin{figure}[htbp]
  \centering
  \includegraphics[width=\linewidth]{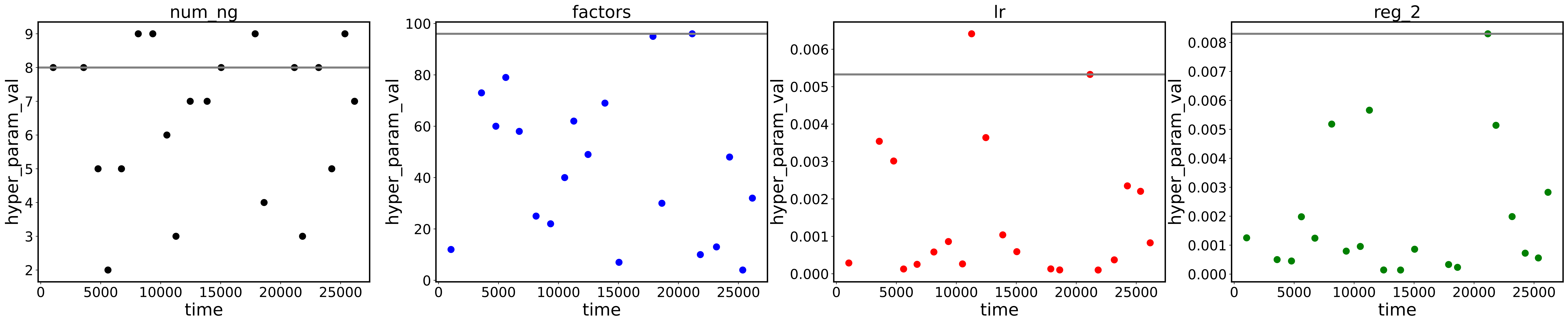}
  \caption{The Search Process of Using the Random Algorithm to Find the Best Hyperparameters of BPRMF on LastFM, and the Gray Solid Line in Each Sub-graph Represents the Best Value of the Corresponding Hyperparameter in the Entire Search Process.}
  \label{fig8}
\end{figure}

\begin{figure}[htbp]
  \centering
  \includegraphics[width=\linewidth]{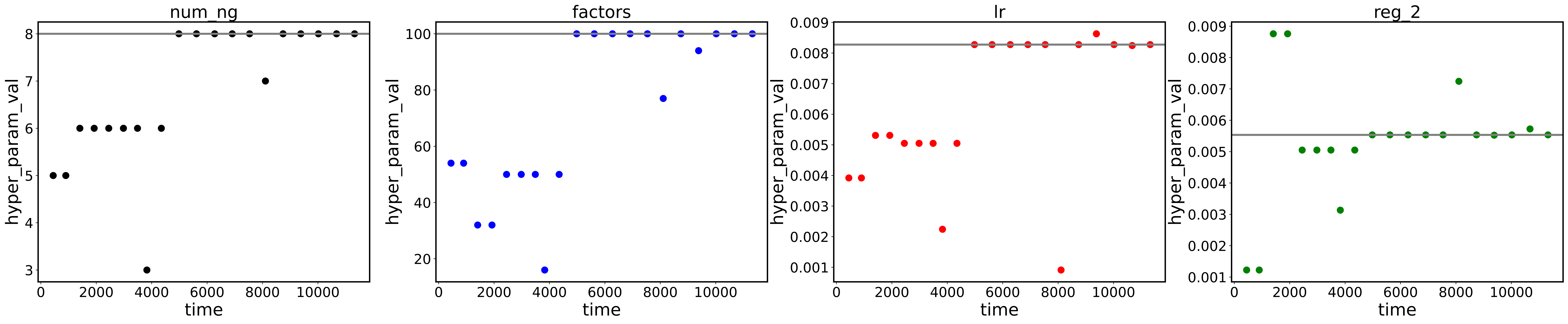}
  \caption{The Search Process of Using the SMAC Algorithm to Find the Best Hyperparameters of BPRMF on LastFM, and the Gray Solid Line in Each Sub-graph Represents the Best Value of the Corresponding Hyperparameter in the Entire Search Process.}
  \label{fig9}
\end{figure}

\begin{figure}[htbp]
  \centering
  \includegraphics[width=\linewidth]{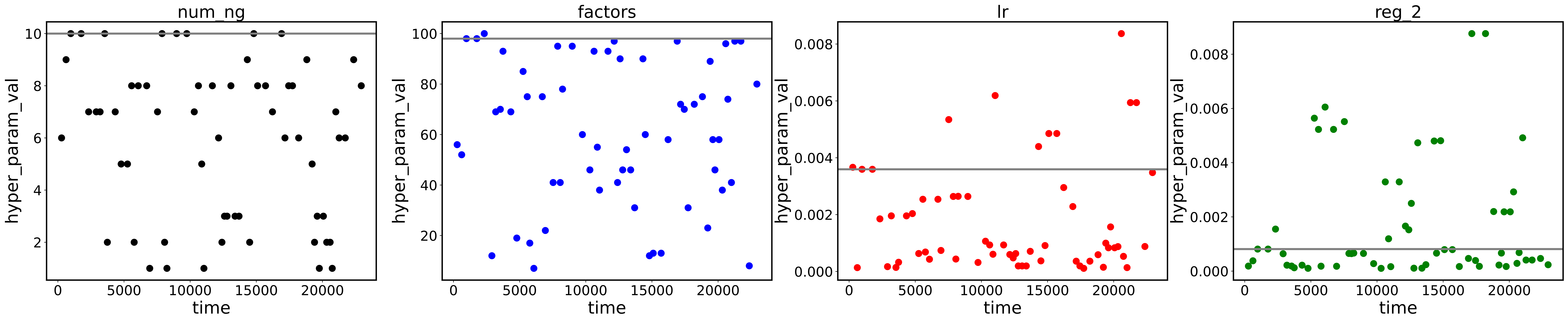}
  \caption{The Search Process of Using the Hyperband Algorithm to Find the Best Hyperparameters of BPRMF on LastFM, and the Gray Solid Line in Each Sub-graph Represents the Best Value of the Corresponding Hyperparameter in the Entire Search Process.}
  \label{fig10}
\end{figure}

\begin{figure}[htbp]
  \centering
  \includegraphics[width=\linewidth]{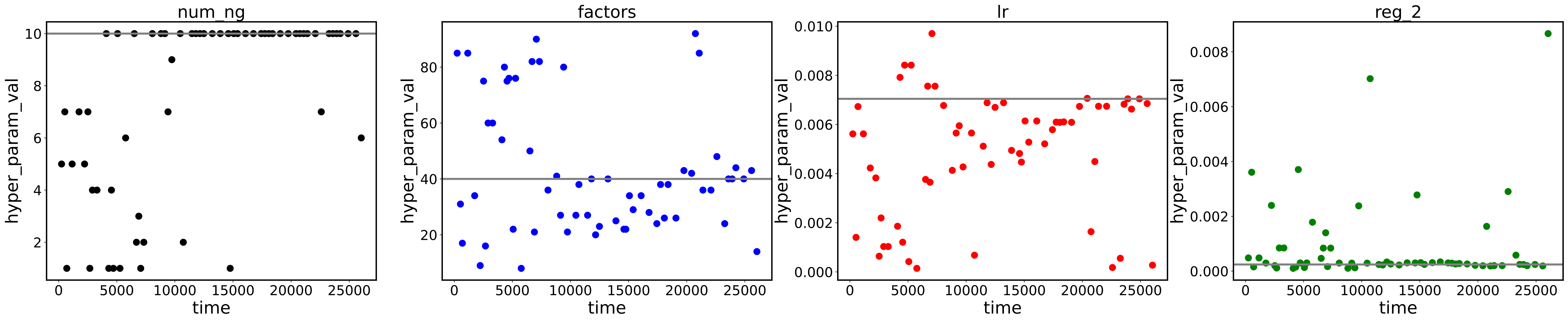}
  \caption{The Search Process of Using the BOHB Algorithm to Find the Best Hyperparameters of BPRMF on LastFM, and the Gray Solid Line in Each Sub-graph Represents the Best Value of the Corresponding Hyperparameter in the Entire Search Process.}
  \label{fig11}
\end{figure}

Figures \ref{fig8}-\ref{fig11} are the selected hyperparameter search process using the Random, SMAC, Hyperband, and BOHB optimization algorithms on the LastFM recommendation scenario with the BPRMF model. The horizontal axis represents the running time, and the vertical axis represents the collected values of each hyperparameter. 
The hyperparameters, namely ``num\_ng'', ``factors'', ``lr'', and ``reg\_2'', are arranged from left to right, and the gray solid line represents the best value found by the corresponding optimization technique. 
Figures \ref{fig8}-\ref{fig11} reveal that the various hyperparameter search algorithms differ significantly in their sampling process, and an analysis of the search process is presented in conjunction with the optimal hyperparameter combination listed in Table \ref{tab12}. The random search algorithm randomly samples hyperparameters, leading to a relatively dispersed distribution of sample points in each hyperparameter subspace. On the other hand, the SMAC algorithm tends to sample areas with better evaluation results as the experiment progresses, resulting in a higher density of sample points in the hyperparameter area with good performance in the later stage of the experiment. The search process of Hyperband and BOHB are shown in Figures \ref{fig10} and \ref{fig11}. Compared with other sampling processes, Hyperband and BOHB algorithms have a higher total sampling point count, indicating that they explore the hyperparameter search space more comprehensively. Notably, in our experimental settings, only $32$ sets of collected hyperparameter combinations in the Hyperband and BOHB algorithms have completed the full $30$ epochs of training, while the rest of the sampled hyperparameter combinations have only partially finished training before evaluation, demonstrating the superiority of these two algorithms in terms of saving computing resources and exploring the search space. Furthermore, an observation of the sampling process of the Hyperband and BOHB algorithms suggests that the BOHB algorithm based on Bayesian thinking yields more concentrated sampling results, while the Hyperband algorithm based on random sampling has a more uniform distribution of sample points.

\section{Conclusions and Future Work}
This paper examined the Top-N implicit recommendation problem and aimed to optimize the benchmark recommendation algorithms typically used in comparative experiments by employing hyperparameter optimization algorithms. The proposed research methodology follows the principles of a fair comparison, using seven types of hyperparameter optimization algorithms to fine-tune six common recommendation algorithms on three datasets. The experimental findings suggest that popular and simple benchmark recommendation algorithms can achieve stable and satisfactory evaluation results by leveraging Anneal and TPE algorithms. However, more complex models like NeuMF and NGCF require Hyperband and BOHB algorithms, which offer higher search efficiency and better recommendation performance. The study provides a fair basis for comparison in model designs in recommender systems based on hyperparameter optimization.

To further investigate the topic, future research should include relevant experiments on all listed datasets and recommendation system algorithms to provide more comprehensive supports for much more solid conclusions. Additionally, the study focuses on three types of hyperparameter optimization algorithms, and future research can explore the fourth type of hyperparameter optimization algorithms based on hypergradient to achieve more comprehensive findings.

\bibliographystyle{ACM-Reference-Format}
\bibliography{acmart}


\begin{thebibliography}{32}


\ifx \showCODEN    \undefined \def \showCODEN     #1{\unskip}     \fi
\ifx \showDOI      \undefined \def \showDOI       #1{#1}\fi
\ifx \showISBNx    \undefined \def \showISBNx     #1{\unskip}     \fi
\ifx \showISBNxiii \undefined \def \showISBNxiii  #1{\unskip}     \fi
\ifx \showISSN     \undefined \def \showISSN      #1{\unskip}     \fi
\ifx \showLCCN     \undefined \def \showLCCN      #1{\unskip}     \fi
\ifx \shownote     \undefined \def \shownote      #1{#1}          \fi
\ifx \showarticletitle \undefined \def \showarticletitle #1{#1}   \fi
\ifx \showURL      \undefined \def \showURL       {\relax}        \fi
\providecommand\bibfield[2]{#2}
\providecommand\bibinfo[2]{#2}
\providecommand\natexlab[1]{#1}
\providecommand\showeprint[2][]{arXiv:#2}

\bibitem[Bengio(2000)]%
        {bengio2000gradient}
\bibfield{author}{\bibinfo{person}{Yoshua Bengio}.}
  \bibinfo{year}{2000}\natexlab{}.
\newblock \showarticletitle{Gradient-based optimization of hyperparameters}.
\newblock \bibinfo{journal}{\emph{Neural Computation}} \bibinfo{volume}{12},
  \bibinfo{number}{8} (\bibinfo{year}{2000}), \bibinfo{pages}{1889--1900}.
\newblock


\bibitem[Bergstra et~al\mbox{.}(2011)]%
        {bergstra2011algorithms}
\bibfield{author}{\bibinfo{person}{James Bergstra}, \bibinfo{person}{R{\'e}mi
  Bardenet}, \bibinfo{person}{Yoshua Bengio}, {and} \bibinfo{person}{Bal{\'a}zs
  K{\'e}gl}.} \bibinfo{year}{2011}\natexlab{}.
\newblock \showarticletitle{Algorithms for hyper-parameter optimization}.
\newblock \bibinfo{journal}{\emph{Advances in Neural Information Processing
  Systems}}  \bibinfo{volume}{24} (\bibinfo{year}{2011}).
\newblock


\bibitem[Bergstra and Bengio(2012)]%
        {bergstra2012random}
\bibfield{author}{\bibinfo{person}{James Bergstra} {and}
  \bibinfo{person}{Yoshua Bengio}.} \bibinfo{year}{2012}\natexlab{}.
\newblock \showarticletitle{Random search for hyper-parameter optimization.}
\newblock \bibinfo{journal}{\emph{Journal of Machine Learning Research}}
  \bibinfo{volume}{13}, \bibinfo{number}{2} (\bibinfo{year}{2012}).
\newblock


\bibitem[Cheng et~al\mbox{.}(2016)]%
        {cheng2016wide}
\bibfield{author}{\bibinfo{person}{Heng-Tze Cheng}, \bibinfo{person}{Levent
  Koc}, \bibinfo{person}{Jeremiah Harmsen}, \bibinfo{person}{Tal Shaked},
  \bibinfo{person}{Tushar Chandra}, \bibinfo{person}{Hrishi Aradhye},
  \bibinfo{person}{Glen Anderson}, \bibinfo{person}{Greg Corrado},
  \bibinfo{person}{Wei Chai}, \bibinfo{person}{Mustafa Ispir}, {et~al\mbox{.}}}
  \bibinfo{year}{2016}\natexlab{}.
\newblock \showarticletitle{Wide \& deep learning for recommender systems}. In
  \bibinfo{booktitle}{\emph{Proceedings of the 1st Workshop on Deep Learning
  for Recommender Systems}}. \bibinfo{pages}{7--10}.
\newblock


\bibitem[Coates et~al\mbox{.}(2011)]%
        {coates2011analysis}
\bibfield{author}{\bibinfo{person}{Adam Coates}, \bibinfo{person}{Andrew Ng},
  {and} \bibinfo{person}{Honglak Lee}.} \bibinfo{year}{2011}\natexlab{}.
\newblock \showarticletitle{An analysis of single-layer networks in
  unsupervised feature learning}. In \bibinfo{booktitle}{\emph{Proceedings of
  the Fourteenth International Conference on Artificial Intelligence and
  Statistics}}. JMLR Workshop and Conference Proceedings,
  \bibinfo{pages}{215--223}.
\newblock


\bibitem[Cremonesi et~al\mbox{.}(2010)]%
        {cremonesi2010performance}
\bibfield{author}{\bibinfo{person}{Paolo Cremonesi}, \bibinfo{person}{Yehuda
  Koren}, {and} \bibinfo{person}{Roberto Turrin}.}
  \bibinfo{year}{2010}\natexlab{}.
\newblock \showarticletitle{Performance of recommender algorithms on top-n
  recommendation tasks}. In \bibinfo{booktitle}{\emph{Proceedings of the Fourth
  ACM Conference on Recommender Systems}}. \bibinfo{pages}{39--46}.
\newblock


\bibitem[Domhan et~al\mbox{.}(2015)]%
        {domhan2015speeding}
\bibfield{author}{\bibinfo{person}{Tobias Domhan}, \bibinfo{person}{Jost~Tobias
  Springenberg}, {and} \bibinfo{person}{Frank Hutter}.}
  \bibinfo{year}{2015}\natexlab{}.
\newblock \showarticletitle{Speeding up automatic hyperparameter optimization
  of deep neural networks by extrapolation of learning curves}. In
  \bibinfo{booktitle}{\emph{Twenty-fourth International Joint Conference on
  Artificial Intelligence}}.
\newblock


\bibitem[Falkner et~al\mbox{.}(2018)]%
        {falkner2018bohb}
\bibfield{author}{\bibinfo{person}{Stefan Falkner}, \bibinfo{person}{Aaron
  Klein}, {and} \bibinfo{person}{Frank Hutter}.}
  \bibinfo{year}{2018}\natexlab{}.
\newblock \showarticletitle{BOHB: Robust and efficient hyperparameter
  optimization at scale}. In \bibinfo{booktitle}{\emph{International Conference
  on Machine Learning}}. PMLR, \bibinfo{pages}{1437--1446}.
\newblock


\bibitem[Franceschi et~al\mbox{.}(2017)]%
        {franceschi2017forward}
\bibfield{author}{\bibinfo{person}{Luca Franceschi}, \bibinfo{person}{Michele
  Donini}, \bibinfo{person}{Paolo Frasconi}, {and}
  \bibinfo{person}{Massimiliano Pontil}.} \bibinfo{year}{2017}\natexlab{}.
\newblock \showarticletitle{Forward and reverse gradient-based hyperparameter
  optimization}. In \bibinfo{booktitle}{\emph{International Conference on
  Machine Learning}}. PMLR, \bibinfo{pages}{1165--1173}.
\newblock


\bibitem[He et~al\mbox{.}(2017)]%
        {he2017neural}
\bibfield{author}{\bibinfo{person}{Xiangnan He}, \bibinfo{person}{Lizi Liao},
  \bibinfo{person}{Hanwang Zhang}, \bibinfo{person}{Liqiang Nie},
  \bibinfo{person}{Xia Hu}, {and} \bibinfo{person}{Tat-Seng Chua}.}
  \bibinfo{year}{2017}\natexlab{}.
\newblock \showarticletitle{Neural collaborative filtering}. In
  \bibinfo{booktitle}{\emph{Proceedings of the 26th International Conference on
  World Wide Web}}. \bibinfo{pages}{173--182}.
\newblock


\bibitem[Hinton(2012)]%
        {hinton2012practical}
\bibfield{author}{\bibinfo{person}{Geoffrey~E Hinton}.}
  \bibinfo{year}{2012}\natexlab{}.
\newblock \showarticletitle{A practical guide to training restricted Boltzmann
  machines}.
\newblock \bibinfo{journal}{\emph{Neural Networks: Tricks of the Trade: Second
  Edition}} (\bibinfo{year}{2012}), \bibinfo{pages}{599--619}.
\newblock


\bibitem[Hu et~al\mbox{.}(2008)]%
        {hu2008collaborative}
\bibfield{author}{\bibinfo{person}{Yifan Hu}, \bibinfo{person}{Yehuda Koren},
  {and} \bibinfo{person}{Chris Volinsky}.} \bibinfo{year}{2008}\natexlab{}.
\newblock \showarticletitle{Collaborative filtering for implicit feedback
  datasets}. In \bibinfo{booktitle}{\emph{2008 Eighth IEEE International
  Conference on Data Mining}}. Ieee, \bibinfo{pages}{263--272}.
\newblock


\bibitem[Hutter et~al\mbox{.}(2011)]%
        {hutter2011sequential}
\bibfield{author}{\bibinfo{person}{Frank Hutter}, \bibinfo{person}{Holger~H
  Hoos}, {and} \bibinfo{person}{Kevin Leyton-Brown}.}
  \bibinfo{year}{2011}\natexlab{}.
\newblock \showarticletitle{Sequential model-based optimization for general
  algorithm configuration}. In \bibinfo{booktitle}{\emph{Learning and
  Intelligent Optimization: 5th International Conference, LION 5, Rome, Italy,
  January 17-21, 2011. Selected Papers 5}}. Springer,
  \bibinfo{pages}{507--523}.
\newblock


\bibitem[Hutter et~al\mbox{.}(2010)]%
        {hutter2010time}
\bibfield{author}{\bibinfo{person}{Frank Hutter}, \bibinfo{person}{Holger~H
  Hoos}, \bibinfo{person}{Kevin Leyton-Brown}, {and} \bibinfo{person}{Kevin
  Murphy}.} \bibinfo{year}{2010}\natexlab{}.
\newblock \showarticletitle{Time-bounded sequential parameter optimization}. In
  \bibinfo{booktitle}{\emph{Learning and Intelligent Optimization: 4th
  International Conference, LION 4, Venice, Italy, January 18-22, 2010.
  Selected Papers 4}}. Springer, \bibinfo{pages}{281--298}.
\newblock


\bibitem[Klein et~al\mbox{.}(2017)]%
        {klein2017learning}
\bibfield{author}{\bibinfo{person}{Aaron Klein}, \bibinfo{person}{Stefan
  Falkner}, \bibinfo{person}{Jost~Tobias Springenberg}, {and}
  \bibinfo{person}{Frank Hutter}.} \bibinfo{year}{2017}\natexlab{}.
\newblock \showarticletitle{Learning curve prediction with Bayesian neural
  networks}. In \bibinfo{booktitle}{\emph{International Conference on Learning
  Representations}}.
\newblock


\bibitem[Koren(2008)]%
        {koren2008factorization}
\bibfield{author}{\bibinfo{person}{Yehuda Koren}.}
  \bibinfo{year}{2008}\natexlab{}.
\newblock \showarticletitle{Factorization meets the neighborhood: a
  multifaceted collaborative filtering model}. In
  \bibinfo{booktitle}{\emph{Proceedings of the 14th ACM SIGKDD International
  Conference on Knowledge Discovery and Data Mining}}.
  \bibinfo{pages}{426--434}.
\newblock


\bibitem[Li et~al\mbox{.}(2017)]%
        {li2017hyperband}
\bibfield{author}{\bibinfo{person}{Lisha Li}, \bibinfo{person}{Kevin Jamieson},
  \bibinfo{person}{Giulia DeSalvo}, \bibinfo{person}{Afshin Rostamizadeh},
  {and} \bibinfo{person}{Ameet Talwalkar}.} \bibinfo{year}{2017}\natexlab{}.
\newblock \showarticletitle{Hyperband: Bandit-based configuration evaluation
  for hyperparameter optimization}. In \bibinfo{booktitle}{\emph{International
  Conference on Learning Representations}}.
\newblock


\bibitem[Lin(2019)]%
        {lin2019neural}
\bibfield{author}{\bibinfo{person}{Jimmy Lin}.}
  \bibinfo{year}{2019}\natexlab{}.
\newblock \showarticletitle{The neural hype and comparisons against weak
  baselines}. In \bibinfo{booktitle}{\emph{ACM SIGIR Forum}},
  Vol.~\bibinfo{volume}{52}. ACM New York, NY, USA, \bibinfo{pages}{40--51}.
\newblock


\bibitem[Lops et~al\mbox{.}(2011)]%
        {lops2011content}
\bibfield{author}{\bibinfo{person}{Pasquale Lops}, \bibinfo{person}{Marco
  De~Gemmis}, {and} \bibinfo{person}{Giovanni Semeraro}.}
  \bibinfo{year}{2011}\natexlab{}.
\newblock \showarticletitle{Content-based recommender systems: State of the art
  and trends}.
\newblock \bibinfo{journal}{\emph{Recommender Systems Handbook}}
  (\bibinfo{year}{2011}), \bibinfo{pages}{73--105}.
\newblock


\bibitem[Ludewig and Jannach(2018)]%
        {ludewig2018evaluation}
\bibfield{author}{\bibinfo{person}{Malte Ludewig} {and}
  \bibinfo{person}{Dietmar Jannach}.} \bibinfo{year}{2018}\natexlab{}.
\newblock \showarticletitle{Evaluation of session-based recommendation
  algorithms}.
\newblock \bibinfo{journal}{\emph{User Modeling and User-Adapted Interaction}}
  \bibinfo{volume}{28} (\bibinfo{year}{2018}), \bibinfo{pages}{331--390}.
\newblock


\bibitem[Pedregosa(2016)]%
        {pedregosa2016hyperparameter}
\bibfield{author}{\bibinfo{person}{Fabian Pedregosa}.}
  \bibinfo{year}{2016}\natexlab{}.
\newblock \showarticletitle{Hyperparameter optimization with approximate
  gradient}. In \bibinfo{booktitle}{\emph{International Conference on Machine
  Learning}}. PMLR, \bibinfo{pages}{737--746}.
\newblock


\bibitem[Pinto et~al\mbox{.}(2009)]%
        {pinto2009high}
\bibfield{author}{\bibinfo{person}{Nicolas Pinto}, \bibinfo{person}{David
  Doukhan}, \bibinfo{person}{James~J DiCarlo}, {and} \bibinfo{person}{David~D
  Cox}.} \bibinfo{year}{2009}\natexlab{}.
\newblock \showarticletitle{A high-throughput screening approach to discovering
  good forms of biologically inspired visual representation}.
\newblock \bibinfo{journal}{\emph{PLoS Computational Biology}}
  \bibinfo{volume}{5}, \bibinfo{number}{11} (\bibinfo{year}{2009}),
  \bibinfo{pages}{e1000579}.
\newblock


\bibitem[Rendle(2010)]%
        {rendle2010factorization}
\bibfield{author}{\bibinfo{person}{Steffen Rendle}.}
  \bibinfo{year}{2010}\natexlab{}.
\newblock \showarticletitle{Factorization machines}. In
  \bibinfo{booktitle}{\emph{2010 IEEE International Conference on Data
  Mining}}. IEEE, \bibinfo{pages}{995--1000}.
\newblock


\bibitem[Rendle et~al\mbox{.}(2012)]%
        {rendle2012bpr}
\bibfield{author}{\bibinfo{person}{Steffen Rendle}, \bibinfo{person}{Christoph
  Freudenthaler}, \bibinfo{person}{Zeno Gantner}, {and} \bibinfo{person}{Lars
  Schmidt-Thieme}.} \bibinfo{year}{2012}\natexlab{}.
\newblock \showarticletitle{BPR: Bayesian personalized ranking from implicit
  feedback}.
\newblock \bibinfo{journal}{\emph{arXiv preprint arXiv:1205.2618}}
  (\bibinfo{year}{2012}).
\newblock


\bibitem[Shi and Gu(2021)]%
        {shi2021improved}
\bibfield{author}{\bibinfo{person}{Wanli Shi} {and} \bibinfo{person}{Bin Gu}.}
  \bibinfo{year}{2021}\natexlab{}.
\newblock \showarticletitle{Improved Penalty Method via Doubly Stochastic
  Gradients for Bilevel Hyperparameter Optimization}. In
  \bibinfo{booktitle}{\emph{Proceedings of the AAAI Conference on Artificial
  Intelligence}}, Vol.~\bibinfo{volume}{35}. \bibinfo{pages}{9621--9629}.
\newblock


\bibitem[Snoek et~al\mbox{.}(2012)]%
        {snoek2012practical}
\bibfield{author}{\bibinfo{person}{Jasper Snoek}, \bibinfo{person}{Hugo
  Larochelle}, {and} \bibinfo{person}{Ryan~P Adams}.}
  \bibinfo{year}{2012}\natexlab{}.
\newblock \showarticletitle{Practical bayesian optimization of machine learning
  algorithms}.
\newblock \bibinfo{journal}{\emph{Advances in Neural Information Processing
  Systems}}  \bibinfo{volume}{25} (\bibinfo{year}{2012}).
\newblock


\bibitem[Sun et~al\mbox{.}(2022)]%
        {sun2022daisyrec}
\bibfield{author}{\bibinfo{person}{Zhu Sun}, \bibinfo{person}{Hui Fang},
  \bibinfo{person}{Jie Yang}, \bibinfo{person}{Xinghua Qu},
  \bibinfo{person}{Hongyang Liu}, \bibinfo{person}{Di Yu},
  \bibinfo{person}{Yew-Soon Ong}, {and} \bibinfo{person}{Jie Zhang}.}
  \bibinfo{year}{2022}\natexlab{}.
\newblock \showarticletitle{DaisyRec 2.0: Benchmarking Recommendation for
  Rigorous Evaluation}.
\newblock \bibinfo{journal}{\emph{IEEE Transactions on Pattern Analysis and
  Machine Intelligence}} (\bibinfo{year}{2022}), \bibinfo{pages}{1--20}.
\newblock
\urldef\tempurl%
\url{https://doi.org/10.1109/TPAMI.2022.3231891}
\showDOI{\tempurl}


\bibitem[Sun et~al\mbox{.}(2019)]%
        {sun2019research}
\bibfield{author}{\bibinfo{person}{Zhu Sun}, \bibinfo{person}{Qing Guo},
  \bibinfo{person}{Jie Yang}, \bibinfo{person}{Hui Fang},
  \bibinfo{person}{Guibing Guo}, \bibinfo{person}{Jie Zhang}, {and}
  \bibinfo{person}{Robin Burke}.} \bibinfo{year}{2019}\natexlab{}.
\newblock \showarticletitle{Research commentary on recommendations with side
  information: A survey and research directions}.
\newblock \bibinfo{journal}{\emph{Electronic Commerce Research and
  Applications}}  \bibinfo{volume}{37} (\bibinfo{year}{2019}),
  \bibinfo{pages}{100879}.
\newblock


\bibitem[Sun et~al\mbox{.}(2020)]%
        {sun2020we}
\bibfield{author}{\bibinfo{person}{Zhu Sun}, \bibinfo{person}{Di Yu},
  \bibinfo{person}{Hui Fang}, \bibinfo{person}{Jie Yang},
  \bibinfo{person}{Xinghua Qu}, \bibinfo{person}{Jie Zhang}, {and}
  \bibinfo{person}{Cong Geng}.} \bibinfo{year}{2020}\natexlab{}.
\newblock \showarticletitle{Are we evaluating rigorously? benchmarking
  recommendation for reproducible evaluation and fair comparison}. In
  \bibinfo{booktitle}{\emph{Proceedings of the 14th ACM Conference on
  Recommender Systems}}. \bibinfo{pages}{23--32}.
\newblock


\bibitem[Van~Laarhoven et~al\mbox{.}(1987)]%
        {van1987simulated}
\bibfield{author}{\bibinfo{person}{Peter~JM Van~Laarhoven},
  \bibinfo{person}{Emile~HL Aarts}, \bibinfo{person}{Peter~JM van Laarhoven},
  {and} \bibinfo{person}{Emile~HL Aarts}.} \bibinfo{year}{1987}\natexlab{}.
\newblock \bibinfo{booktitle}{\emph{Simulated annealing}}.
\newblock \bibinfo{publisher}{Springer}.
\newblock


\bibitem[Wang et~al\mbox{.}(2019)]%
        {wang2019neural}
\bibfield{author}{\bibinfo{person}{Xiang Wang}, \bibinfo{person}{Xiangnan He},
  \bibinfo{person}{Meng Wang}, \bibinfo{person}{Fuli Feng}, {and}
  \bibinfo{person}{Tat-Seng Chua}.} \bibinfo{year}{2019}\natexlab{}.
\newblock \showarticletitle{Neural graph collaborative filtering}. In
  \bibinfo{booktitle}{\emph{Proceedings of the 42nd international ACM SIGIR
  conference on Research and development in Information Retrieval}}.
  \bibinfo{pages}{165--174}.
\newblock


\bibitem[Wang et~al\mbox{.}(2016)]%
        {wang2016bayesian}
\bibfield{author}{\bibinfo{person}{Ziyu Wang}, \bibinfo{person}{Frank Hutter},
  \bibinfo{person}{Masrour Zoghi}, \bibinfo{person}{David Matheson}, {and}
  \bibinfo{person}{Nando De~Feitas}.} \bibinfo{year}{2016}\natexlab{}.
\newblock \showarticletitle{Bayesian optimization in a billion dimensions via
  random embeddings}.
\newblock \bibinfo{journal}{\emph{Journal of Artificial Intelligence Research}}
   \bibinfo{volume}{55} (\bibinfo{year}{2016}), \bibinfo{pages}{361--387}.
\newblock


\end{thebibliography}
\end{document}